\documentclass{article}
\usepackage{graphicx}

\newcommand{\eofi}{M}

\begin{document}
\title{Nonparametric modeling and spatiotemporal dynamical systems}
\author{Markus Abel
\vspace{1cm}\\
Universit\"at Potsdam, Institut f\"ur Physik, \\Postfach 601553,
D-14415 Potsdam, Germany.
}  
\maketitle

\begin{abstract}  
In this article, it is described how to use statistical data analysis
to obtain models directly from data. The focus is put on finding
nonlinearities within a generalized additive model. These models are
found by the means of backfitting algorithms or more general versions,
like the alternating conditional expectation value method. The method
is illustrated by numerically generated data. As an application the
example of vortex ripple dynamics, a highly complex fluid-granular
system is treated.
\end{abstract} 
\maketitle

 
\section{Introduction}
One of the major goals of spatiotemporal data analysis consists in
inferring a model from a given spatiotemporal data set. Conventional
ways of modeling are purely theoretical arguing and a posteriori
evaluation of possible models with suitable measures. In this article,
statistical methods and applications to infer generalized additive
models from data are presented. The attention is put on
spatio-temporal systems, but there are certainly more possibilities of
using the ideas in physics and engineering.  Generalized additive
models have been introduced to statistical analysis in the 1980's
\cite{Hastie-Tibshirani-86,Hastie-Tibshirani-90}.  The
development of solving algorithms and mathematical proofs has been
developed contemporarily (see \cite{Hastie-Tibshirani-90} and
references therein).

During the last two decades, natural sciences profited a lot by the
increasing computer power of modern machines. In nonlinear and
statistical sciences, old techniques could be exploited to their full
extent and new methods have been developed. The identification of
nonlinearities is one of the important topics in data analysis
of. e.g., chaotic or pattern forming systems
\cite{Kantz-Schreiber-97,Ott-94}. This in turn is feasible by
statistical tools.

Careful application of algorithms and thorough interpretation of
results leads to a direct way to obtain models from data.  Hereafter, 
the basic ideas of the techniques to solve for a generalized additive
model are explained and an application to spatio-temporal
data is given. The intention is not to give a full overview of statistical
methods, nor to report an experiment in detail. Rather, it is 
shown how a consistent procedure to infer models from data can be used
as a feedback to theory which in turn can motivate experimental
designs to yield new data and so on.

The paper is structured as follows: In Sec.~\ref{sec:backfitting} the
algorithmic solutions for generalized additive models are pointed out.
This is done by the example of ACE, the {\bf A}lternating {\bf C}onditional
{\bf E}xpectation value algorithm, which uses the backfitting
technique. This presentation is given some room; even though nothing
new for statisticians, technical points need some explanation to be
understood.  Sec.~\ref{sec:more} provides more information to be used
in the context of spatio-temporal data analysis. In
Sec.~\ref{sec:apps} problems arising from preprocessing of data are
discussed along with the pattern formation example of vortex ripple
dynamics. The paper concludes with a short discussion in
Sec.~\ref{sec:conclusion}.

\section{Generalized additive models and backfitting}
\label{sec:backfitting}
The standard tool of data analysis for physicists and engineers is the
multivariate linear regression, where one fits the model
\begin{equation}
U_0 = C + \sum_{i=1}^{\eofi} \alpha_i U_i + \epsilon\;
\end{equation}
to experimental data. The symbols $\epsilon$ and $U_i$ (upper case
letters), $i=0,\dots,\eofi$ denote a random process with given
distribution, $C$ is a constant and the $\alpha_i$ are model
parameters. One measures a realization $u_i$ (lower case letters),
$i=0,\dots,\eofi$ of the random process and finds estimates
$\hat{\alpha_i}$ for the parameters, e.g., by least squares fitting
\cite{NR}. Throughout the text, estimations are denoted by a hat,
only where ambiguity exists, otherwise additional symbols are omitted.

There are many ways of generalization. One often used way is
the generalized linear regression model,
\begin{equation}
U_0 = C + \sum_{i=1}^{\eofi} \alpha_i f_i(U_i) + \epsilon\;.
\end{equation}
In this model, the data analyst chooses beforehand the functions $f_i$
due to some prior knowledge about the system or simply by
guessing. Then, a linear regression procedure yields estimates for the
parameters $\alpha_i$.  So, the model is expressed in parametric form
by the estimates $\hat{\alpha_i}$. For details about linear
regression, see, e.g., \cite{NR,Montgomery-92}.

If no obvious choice for a function is at hand, it is desirable to
fit the nonparametric model
\begin{equation}
\label{eq:gam}
U_0 = C +\sum_{i=1}^{\eofi} f_i(U_i) + \epsilon\;.
\end{equation}
Given some data, estimates $\hat{f_i}$ must result.  With this kind of
modeling no prior knowledge about functional dependences is needed as
input; one can find very general functional forms. A disadvantage is
the absence of even more general terms, like $f_i(U_i,\,U_j)$, $j\neq
i$. The reason for this lack is a practical one: as will become clear
below, any implementation finding functions of $d$ variables has to
compete with the curse of dimensionality.  For the generalized
additive model~(\ref{eq:gam}), estimates for the functions $f_i$ are
found by the {\it backfitting algorithm}
\cite{Hastie-Tibshirani-86,Hastie-Tibshirani-90,Haerdle-Hall-93}. This
is an iterative procedure, working by the following rules:
\begin{enumerate}
\item Initialization: $C=E(U_0)$, $f_i=f_i^0$, $i=1,...,\eofi$.
\item Iteration:  for $i=1,...,\eofi$ calculate
\[ f_i=E\left(U_0-C-\sum_{k\neq i} f_k(U_k) | U_k  \right) \]
until convergence,
\end{enumerate}
where $E()$ denotes the expectation value and the $f_i^0$ are some
appropriate initial settings. In applications, one has to obtain the
functions from data as realizations of the generating process. Then,
this operator has to be replaced by its estimator.  In each iteration
step we estimate the functions by $\hat{f_i}(U_i) = S_i(\cdot|U_i)$, a
smoothing operator which returns a function dependent on
$U_i$.   

The discussion of properties and choice of the smoother is a subtle
issue and lies far beyond the scope of this paper, a detailed
discussion is given in \cite{Buja-Hastie-Tibshirani-89}.  The examples
presented in this article have been obtained using the simple running
mean smoother (moving average). This is not the best choice for many
problems, e.g., points at the boundaries have to be considered with
care or must be cut - throughout this article, there have been enough
data to afford for the luxury to cut the boundaries, in
spatio-temporal systems there are often many thousand of data points.
Smoothing splines are good smoothers in many situations due to their
differentiability properties
\cite{Hastie-Tibshirani-90,Buja-Hastie-Tibshirani-89}. In any case, a
 careful check should be undertaken for a concrete set of data.

The iteration works by adjusting only for one function, subtracting
all the others from the estimation of $U_0$, i.e. one smoothes the
partial residual $(U_0-\sum_k f_k)$ against $U_i$.  In spatiotemporal
data analysis, on is faced with the task of finding equations of
motion from data. These can be coupled map lattices, a set of ordinary
coupled differential equations, partial differential equations or even
more complicated models. In many of these formulae, the lhs. is linear
(e.g. a time derivative) and Eq.~(\ref{eq:gam}) constitutes an
appropriate additive, nonlinear model.  If the measured data are
nonlinear transforms through a measuring function, it can be desirable
to model a nonlinear lhs, too:
\begin{equation}
\label{eq:ace1}
f_0(U_0) = C +\sum_{i=1}^{\eofi} f_i(U_i) + \epsilon\;.
\end{equation}
The constant can be absorbed w.l.o.g. into the functions $f_i$.  Now,
the model is more symmetrical and indeed, in physics, often situations
without a clear predictor (lhs) - response (rhs) structure are
found. A typical example is pattern formation where one is interested
in the nonlinear interaction of stationary states. A priori it is not
clear which variables constitute the state space nor is it always
possible to measure them. Data analysis provides then an estimation of
the model as a nonlinear, possibly noninvertible, transformation of
the measured data. As an example, take the relation $U_0=G(U_1+U_2)$
with $G$ some nonlinear invertible function. The inverted equation
$G^{-1}(U_0)=U_1+U_2$ cannot be found with (\ref{eq:gam}), but with
the model (\ref{eq:ace}). An example for this case is given below.

An algorithm to solve (\ref{eq:ace1})  is  ACE,  (the Alternating
Conditional Expectation value algorithm) \cite{BF-85},
it works by  minimizing the squared error
\begin{equation}
\label{eq:ace}
E\left[ f_0(U_0)-\sum_{i=1}^{\eofi} f_i(U_i) \right]^2 \;.
\end{equation}
The method shall be explained stepping from the one-dimensional
estimation over two dimensional to the  M-dimensional problem.
\subsubsection*{1) $U_0 =  f_1(U_1) + \epsilon$.}
\label{eq:reg1}
\begin{figure}
\begin{center}
\includegraphics[clip,angle=0, width=0.8\textwidth]{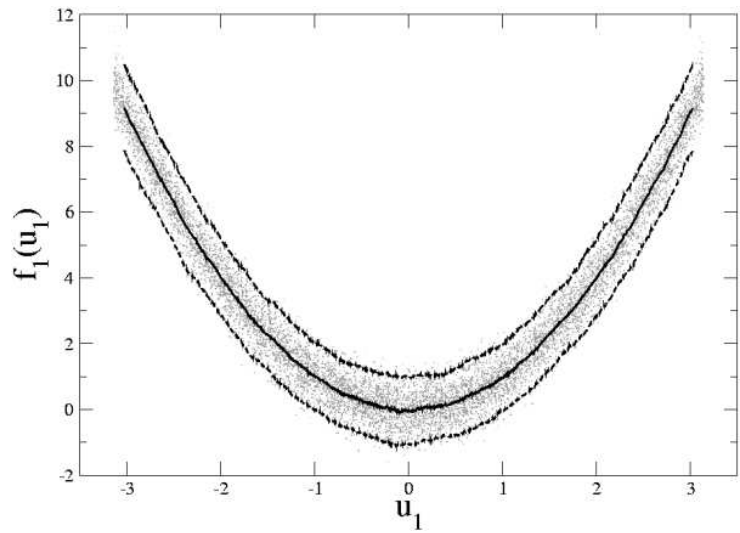}
\caption{One step transformation in the case $f_0=U_0$,
$f_1(U_1)=U_1^2$.  The scatterplot of points $(u_0,u_1)$ is shown by
grey dots (not labelled).  Applying the running mean smoother yields an
estimation for the function $f_1$ (middle, thick black line), above
and below is plotted twice the pointwise standard deviation (upper and
lower black lines).}
\label{fig:ex1}
\end{center}
\end{figure}

For this simple model, one obtains from Eq.~(\ref{eq:ace}) 
\begin{equation}
\label{eq:first}
E\left[ U_0 - f_1(U_1) \right]^2 = min, 
\end{equation}
where minimization is achieved by variation of $f_1$ in the space of
measurable functions \cite{BF-85}. After some elementary
steps one finds the solution \cite{Honerkamp-94}
\begin{equation}
\label{eq:cev1}
f_1(U_1) = E(U_0 |U_1)\;.
\end{equation}
For applications, again a smoother is used as estimator of the
expectation value operator. The result of the smoothing on a set of
numerically produced data is shown for the example $U_0=U_1^2$ in
Fig.~\ref{fig:ex1}. The data has been generated by first drawing 10
000 equally distributed random numbers for $u_1$ in the interval
$(-3,3)$. These numbers have been squared to yield the values for
$u_0$, after this Gaussian noise $N(0,0.1^2)$ has been added. The
bandwidth of the running mean smoother has been set here and below to
200 points.  In Fig.~\ref{fig:ex1}, the scatterplot of the points
$(u_1,u_0)$ is shown (grey) together with the estimate for the function
$f_1$, found by Eq.~(\ref{eq:cev1}) (black, thick line).  As
indication for the estimation error the pointwise standard deviation
has been calculated from the estimated function and the residuals
(upper and lower black lines).

\subsubsection*{2) $\;\; f_0(U_0) =  f_1(U_1) + \epsilon \;$.}
\label{eq:reg2}
In this case, the solution is found iteratively:
\begin{enumerate}
\item Initialization: $f_0(U_0) = \left[ U_0-E(U_0)\right]/\sqrt{var(U_0)}$.
\item Computation of  $f_1(U_1)=E[f_0(U_0)|U_1]$
\item Computation of  $\tilde{f_0}(U_0)=E[f_1(U_1)|U_0]$
\item Normalization:  $f_0(U_0)=\tilde{f_0}(U_0)/\sqrt{var(\tilde{f_0}(U_0))}$.
\item Iteration of  2)-4) until convergence.
\end{enumerate}
A smoother with nonzero bandwidth is contracting and thus the trivial
solution is excluded by normalization (point 4) with the variance $var(\tilde{f_0}(U_0))$.

In order to show how ACE handles non-invertible relations, data ($A
\cos \theta$, $A \sin \theta$) have been generated, with
$\theta=\varphi+\epsilon_\varphi$ and $\varphi$ an equally distributed
random number in $[0,2\pi)$ and $A=1+\epsilon_A$. The noise
$\epsilon_\varphi\,, \epsilon_A$ is Gaussian distributed, with
$N(0,0.05^2)$. This corresponds to the noisy measurement of an
amplitude and a phase with a nonlinear, noninvertible measurement
transformation.

Without noise, one has $U_0= \cos( \varphi)$, $U_1=\sin( \varphi)$ and
thus $U_0^2=1-U_1^2$. With noise, the relation reads to first order
$U_0^2-2\epsilon_A\, U_0=1-U_1^2+2\epsilon_A\, U_1$. This means that
multiplicative noise enters in predictor and response, touching the
errors in variables problem which is relevant for many applications.
The strong, multiplicative transformation of the noise is mirrored
directly in the results.

\begin{figure}
\begin{center}
\includegraphics[clip,angle=0, width=1.0\textwidth]{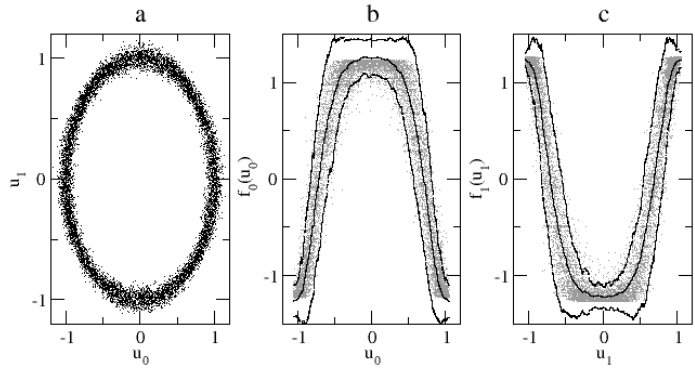}
\vspace{0.5cm}
\caption{Example for $U_0=A\,\sin \theta$, $U_1=A\,\cos \theta$, where
$A$ and $\theta$ include measurement noise (a). The expected
functional relation is noninvertible: $U_0^2+U_1^2=1$, noise on
$A,\theta$ is transformed nonlinearly. The grey dots in (b), (c) show
this effect in the residuals. This in turn influences the result in
that it induces bending at the ends and in the center part of the
results of ACE, where the transformation has its biggest effect $f_0$
and $f_1$ (see the middle line in (b), (c)). The symmetric pointwise
standard deviation is not a good indicator for the asymmetry (upper
and lower lines in (b), (c)), according asymmetric measures should be
calculated.  Functions are shifted to zero mean.}
\label{fig:ex2}
\end{center}
\end{figure}

In Fig.~\ref{fig:ex2} a) the data are displayed in a scatterplot, in
Fig.~\ref{fig:ex2} b) and c), the resulting functions are shown on
top of the respective residuals together with twice the pointwise
standard deviation as an error indicator.  Due to the nonlinear noise
transformation the functions are distorted. The pointwise standard
deviation is less significant as error indicator due to the very
asymmetric local distribution of values, to be seen in the upper and
lower curves of Fig.~\ref{fig:ex2} b) and c). In this case asymmetric
measures should be calculated. The above example  underlines a
conceptual problem when one is faced with errors in the measurement
variables. The general treatment of this problem is not simple, for
more details see \cite{Fan-Truong-93,Carroll-Maca-Ruppert-99} and
references therein.

Even though the precise estimation of the functional dependence is not
possible in the above case, one finds an approximation which would not
be so
bad for many experiments. Please note that linear tools would not be
able to find this relation easily, rather one should undergo a
trial-and-error procedure until a guess for the right form of the
functions is found. Some problems with a non-Gaussian distribution,
however, remain in linear regression, too.

\subsubsection*{3)$\;\; f_0(U_0) = \sum_{i=1}^{\eofi} f_i(U_i) + \epsilon\;$}
\label{eq:reg3}

This equation defines a (noisy) hypersurface in $\eofi$-dimensional space
(in two dimensions, one finds a line). The functions $f_i$,
fall on a hyper{\it plane}: the algorithm
performs a transformation of the points $\{ U_1,\dots,U_{\eofi}\}$ to
the points $\{ f_1,\dots,f_{\eofi}\}$ such that the best plane in the
mean squared sense is found.  

The modification in the algorithm has to be done in the steps 2. and
3.  above. They are replaced by
\begin{enumerate}
\item[2.] Computation of $f_i(U_i)=E[f_0(U_0)-\sum_{k\neq i}f_k(U_k)|U_i]$
by backfitting.
\item[3.] Computation of  $\tilde{f_0}(U_0)=E[\sum_{i}f_i(U_i)\,|\,U_0]$
\end{enumerate}
This shows that the ACE algorithm is based on backfitting.  To
demonstrate a higher-dimensional example, the relation
$U_0=\exp(U_1+U_2^2)$ has been chosen. The variables have been
generated using equally distributed values for $U_1\,,U_2$ in the
interval $(-2,2)$. From the resulting numbers, $u_0$ has been
calculated, thus the smallest possible value of $u_0$ is
$\exp(-2)\simeq 0.14$, the largest is $\exp(6)\simeq 400$. After this
procedure (in contrast to the previous example), Gaussian,
$N(0,0.2^2)$, noise has been added to $u_i$, $i=0,1,2$. This avoids
the problems with transformation of measurement noise, discussed
above. But at this place, another interesting question for the modeler
is raised instead: what happens if one tries to include some
additionally measured variable in the model?

As an example, one can imagine a measurement of four 
variables, where it is not clear a priori which variables are needed
in a minimal model. As a first guess, a generalized additive model
$f_0(U_0)=f_1(U_1)+f_2(U_2)+f_2(U_3)$ will be tested.  This situation
shall be now related to the data generated above: an  additional, equally
distributed variable in the interval $(0,100)$, uncorrelated to
$U_0,\,U_1,\,U_2$, is generated and fed as fourth variable into ACE.

The three-variable additive model is $f_0(U_0)=f_1(U_1)+f_2(U_2)$ and
one expects ACE to find the estimates
$\hat{f_0}(U_0)=\ln(U_0)\,,\hat{f_1}(U_1)=U_1\,,\hat{f_2}(U_2)=U_2^2$
due to the additive structure. For the four dimensional model, one
expects an unchanged result for $\hat{f_0},\,\hat{f_1},\,\hat{f_2}$
plus a zero function $\hat{f_3}(U_3)=0$ in the additional variable.

\begin{figure}
\begin{center}
\includegraphics[clip,angle=0, width=0.75\textwidth]{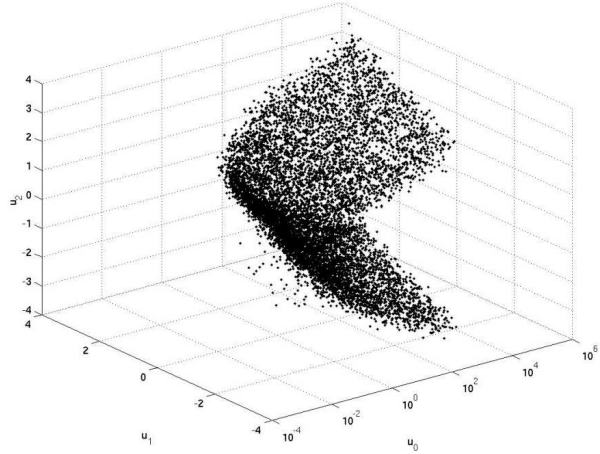}
\caption{Three dimensional plot of the data from
$U_0=\exp(U_1+U_2^2)+\epsilon$.}
\label{fig:ex3-3D}
\end{center}
\end{figure}

\begin{figure}
\begin{center}
\includegraphics[clip,angle=0, width=0.8\textwidth]{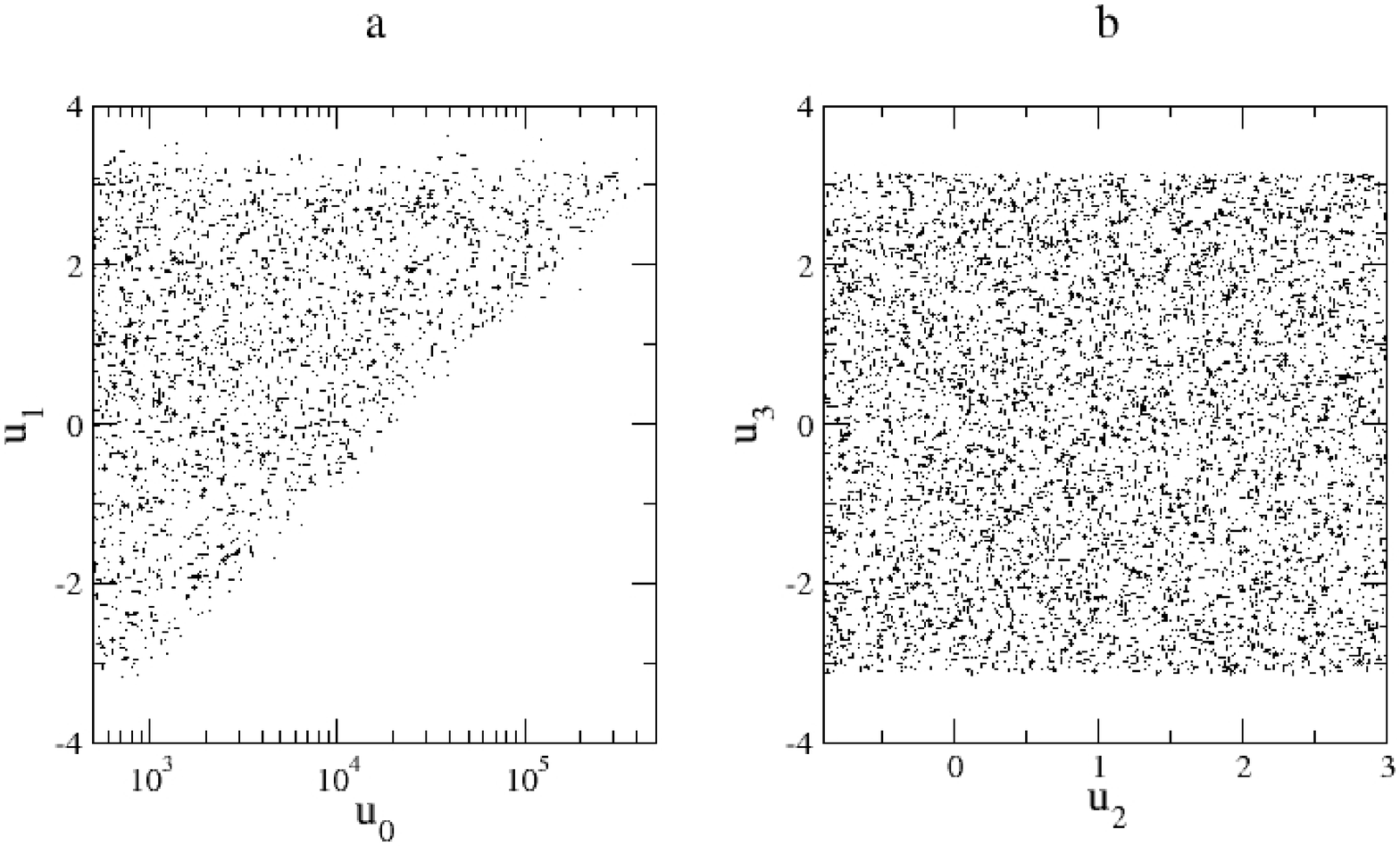}
\caption{Two dimensional scatter plots of a part of the data as
projection on the plane. The examples $u_0\,,u_1$ and $u_2\,,u_3$ are
shown. even ``guessing'' a logarithmic relation for $u_0$, no clear
relation can be read on the left, on the right, the data are
completely random according to the zero correlation of $u_3$ with the
rest.}
\label{fig:ex31}
\end{center}
\end{figure}

\begin{figure}
\begin{center}
\includegraphics[clip,angle=0, width=0.8\textwidth]{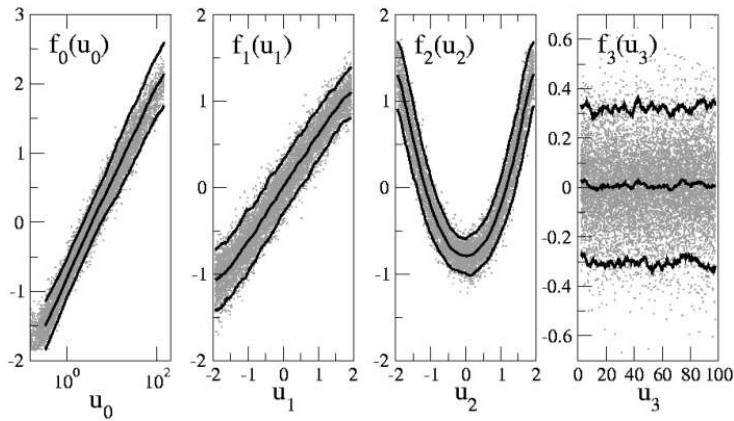}
\caption{Transformations for the given data with twice the pointwise
standard errors (black lines) on top of the residuals (grey). The
logarithm $f_0(u_0)$ is found over three decades to hit numerical
restrictions at the smallest points.  The linear and square functions
$f_1$ and $f_2$ are found very well. The variable $u_3$ is found to be
``unimportant'' with $f_3(u_3)=0$.}
\label{fig:ex32}
\end{center}
\end{figure}

In Fig.~\ref{fig:ex3-3D}, the 3D scatterplot of the generated relevant
data is shown by the points $(\ln(U_0),U_1,U_2)$, the $U_0$-axis is
logarithmic.  A two dimensional representation is not able to display
the full relation, if one plots, e.g., $U_0$ vs. $U_1$
(Fig.~\ref{fig:ex31}), no functional dependence is cognizable. The
transformations $(\hat{f_0},\hat{f_1},\hat{f_2},\hat{f_3})$ found by
ACE are shown in Fig.~\ref{fig:ex32} as functions of their
arguments. The function $f_0$ is found to be logarithmic, 
$f_0=0.5\,\ln(u_0)+c$, $f_1$ is linear with a
slope of $0.5$, and $f_2=a u_2^2+c$ is quadratic also with a factor
$a=0.5$. The factor $0.5$ reflects possible ambiguities in the result,
as well as the addition of the constant, see Sec.~\ref{ssec:pec-ace}.

After multiplying the results by $2$ and adding a constant to $f_0$
and $f_2$ one can speak of a good coincidence of expected and found
result. The additional variable $u_3$ has no dependence on the others 
and yields $f_3=O(10^{-2}) \simeq 0$. Thus, the method is able to identify
uncorrelated terms. A comparison of runs with and without the
additional third term on the rhs showed that the functions are
virtually unchanged, confirming the stability of the algorithm.
Despite the seemingly convincing result it must be noted that results
can be misleading if there exist correlations between an additional
term and ``true'' variables. If the task is to find a minimal model,
one has to consider this effect. The connection of correlation with
ACE is discussed in Sec.\ref{ssec:mc}.

\section{Some further topics}
\label{sec:more}

\subsection{Smoothers and the bias-variance dilemma }
As mentioned above, the choice of the smoother can be
delicate. Typically, to each smoother belongs a smoothing parameter
which characterizes its bandwidth (window width).   
For the running mean smoother, one obtains for problem (\ref{eq:cev1})
\begin{equation}
\hat{f}_1(U_{1,l}) = 1/N_l \sum_{j\in B(U_{0,l})} U_{0,j}\;,
\end{equation}
where the hat denotes estimation and $ B(U_{0,l})$ is a neighborhood
of the point $U_{0,l}$. Increasing bandwidth enlarges
the neighborhood and thus the number $N_l$ of data points in
$B$.  The respective expectation and variance are
\begin{eqnarray}
E(\hat{f}_1(U_{1,l})) &=& 1/N_l \sum_{j\in B(U_{0,l})} f(U_{0,j})\;,\\
var(\hat{f}_1(U_{1,l})) &=& \sigma^2/N_l\;,
\end{eqnarray}
where  $\sigma_l=\sigma$ is assumed for simplicity.  Thus, the
variance shrinks with increasing bandwidth; the bias, however,
increases because more terms different from $f(U_{0,l})$ are
involved. If one wants to find the optimum choice of the bandwidth in order
to have neither large bias nor large variance, one uses measures like
the mean-squared error or the average predictive squared error; these
are evaluated using cross-validation and yield a criterion for the
bandwidth to be chosen. For more information it is  referred to 
the literature \cite{Hastie-Tibshirani-90,Honerkamp-94}.

The iteration procedure should run until convergence.  In fact,
convergence is not guaranteed for all classes of smoothers, asymptotic
properties and convergence is discussed in a rigorous way in
\cite{BF-85,Buja-Hastie-Tibshirani-89,Mammen-99}.  Intuitively, it is
clear that the iteration should stop when the errors on the functions
fall below the noise $\epsilon$ in the model (\ref{eq:ace1}). Then it
is a question how to find properly the respective errors. There are
different ways to estimate the error consistently, depending on the
smoother and the error models used.  Locally, one can choose twice the
pointwise standard error as a characterization, global confidence
criteria are harder to derive
\cite{Hastie-Tibshirani-90,Fan-Truong-93}, an elegant and modern
approach is given by the Bayesian formulation of the backfitting
algorithm \cite{Hastie-Tibshirani-00}. In the case of errors in
variables, care has to be taken to calculate the error correctly, a
Bayesian approach is appealing in that case.  In the examples, the
most naive criterion, i.e., twice the pointwise standard error has
been used.

\subsection{Peculiarities of ACE}
\label{ssec:pec-ace}
The ACE algorithm can be related to other algorithms, namely canonical
variates and alternating least squares methods. Special properties of
all three methods are discussed in \cite{Buja-90}, peculiarities of ACE
are discussed in \cite{Hastie-Tibshirani-90,BF-85}, here a few items shall
be given without  a claim for completeness.\\
i) ACE is symmetric in the predictor and the response, which is quite
unusual for a regression tool, but suits well several setups in
physical applications.\\
ii) Transformations of the variables are not reproduced by ACE. Take, e.g.,
 the case $U_0=f_1(U_1) + \epsilon$. If one applies a
transformation to each of the sides, ACE does not necessarily find
$f_1$ again. This reflects parts of the nature of the inverse problem,
which is in general not uniquely solvable. Especially, adding a constant
to either of the sides in the model is ambiguous in ACE.\\
iii) Different distributions in the model constituents can ``distort'' the resulting
functions. E.g., it may happen that one term in a model with two
variables is Gaussian distributed and another one is equally
distributed in some interval. Then, the Gaussian variable is bent at
the ends (cf. Fig.~\ref{fig:ex32}).

\subsection{Maximal Correlation}
\label{ssec:mc}
The ACE algorithm can be regarded as a regression tool, but there
exists another possible interpretation. Instead of solving the
least squares error minimization problem
\begin{equation}
E[f_0(U_0) - \sum_{i=1}^{\eofi} f_i(U_i)]^2=min 
\end{equation}
one can reformulate the above to 
\begin{equation}\label{eq:mc}
\Psi = corr\left(f_0(U_0),\, \sum_{i=1}^\eofi f_i(U_i)   \right) = max\;,
\end{equation}
where $corr$ denotes the correlation function.  This reflects the
principle of the maximal correlation
\cite{Renyi-59,Bell-62,Csaki-63,Renyi-70}. The functions that fulfill 
Eq.~(\ref{eq:mc}) are called optimal (in the sense of correlation)
transformations, found by the ACE procedure. To have a global measure
for the importance of a single term $f_i(U_i)$ , there are several
possibilities, e.g., one can choose the overall variance, $var(f_i)$,
but this does not reflect correlation with other terms and a noise
term with large variance would be judged to be important. Here, the author
suggests to use Eq.~(\ref{eq:mc}) in a symmetric way,
\begin{equation}
\Psi_i = \left| \frac{ E(f_i\cdot \sum_{j\neq i} f_j)}
{\sqrt{E(f_i^2)\cdot E(\sum_{j\neq i} f_j)^2}}   \right| \;,
\end{equation}
to characterize the importance of a single term.  The use of this
measure is demonstrated for the data from the example used for
Fig.~\ref{fig:ex32}. The  values $\Psi_0=0.9779$,
$\Psi_1=0.9272$, $\Psi_2=0.9690$ and $\Psi_3=0.0426$ are found,
confirming again the unimportance of the additional uncorrelated
variable, this time by the global measure $\Psi_3$. 

If however, the additional variable is correlated, e.g., by passively
following another variable, a high correlation will follow, yielding a
model with more components. In the case of spatiotemporal modeling one
has to be specially careful, because one is  interested in
minimal (in the number of terms) models, but dynamical dependencies
can result easily in correlations, like in the case of slaved variables.

\subsection{Generalizations}
\label{ssec:gen}
One can generalize the procedure to include more complex, non-additive
couplings, e.g. in the three dimensional case, one considers the model
\begin{equation}
\label{eq:general}
U_0 = F(U_1,U_2) + \epsilon\;,
\end{equation}
with $F$ some function, describing a two dimensional surface in a
three dimensional space spanned by $U_0,\,U_1,\,U_2$. Like for
Eq.~(\ref{eq:cev1}), one finds
\begin{equation}
F(U_1,U_2) = E(U_0 |U_1,U_2)\;.
\end{equation}
Going to higher dimensions requires in general more data due to the
curse of dimensionality, additionally one has to decide again which
estimator to use for correct results.

\begin{figure}
\begin{center}
\hfill
\includegraphics[angle=0,width=0.42\textwidth]{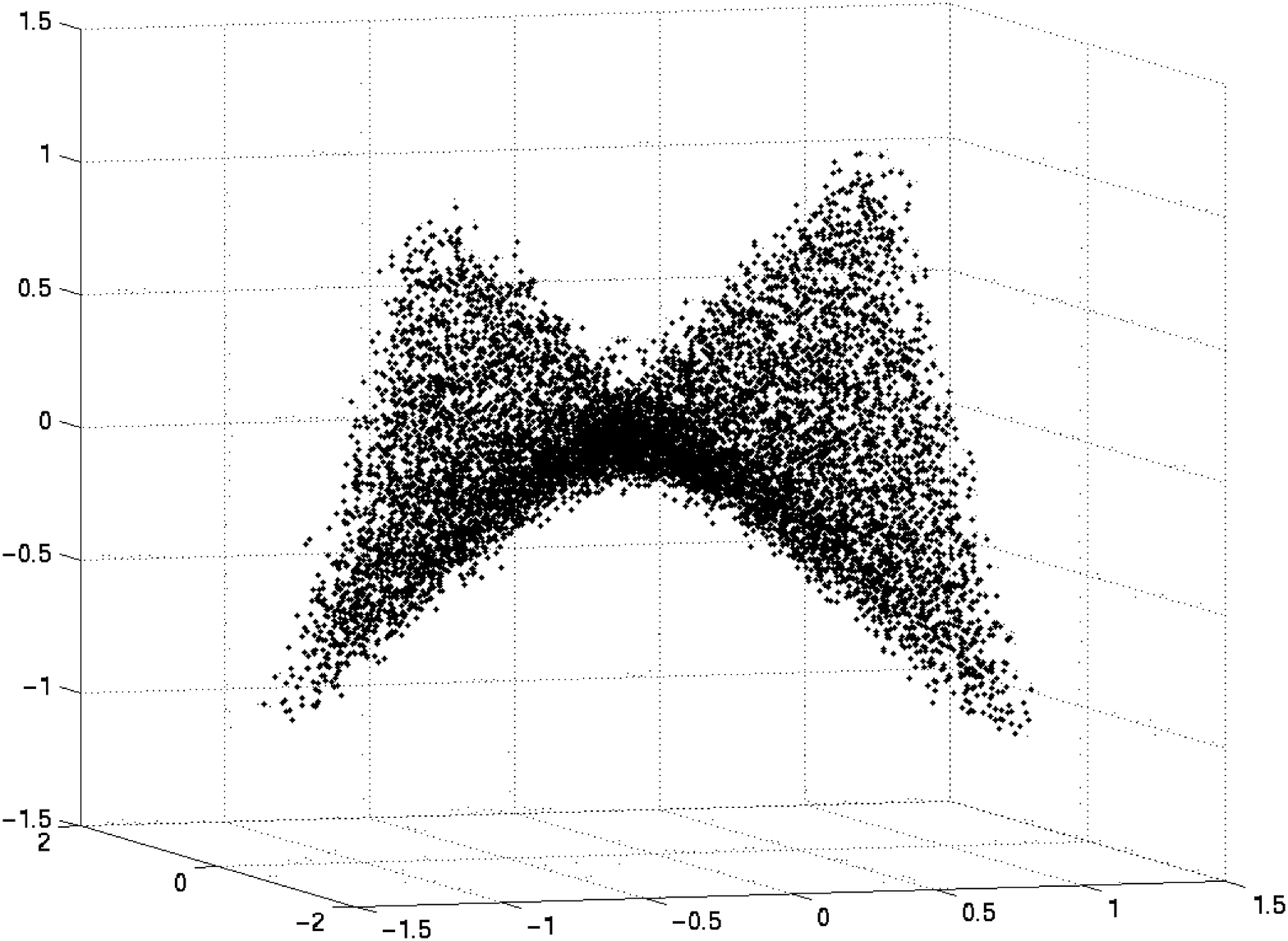}
\hfill
\includegraphics[angle=0,width=0.42\textwidth]{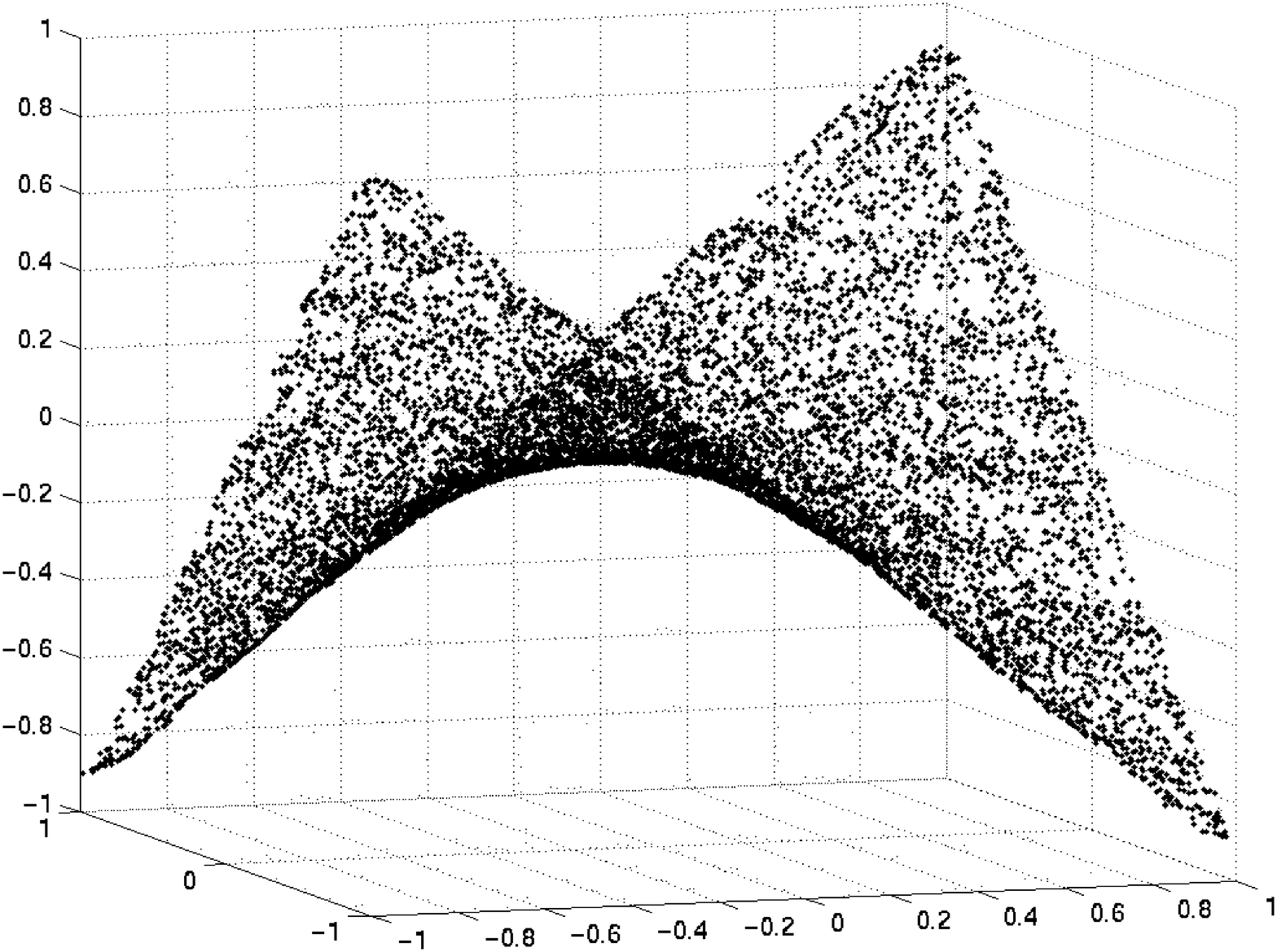}
\hfill
\caption{Left: noisy data for the surface $U_0=U_1\cdot U_2$,
$\sigma=0.05$ for all $U_i$. Right: Optimal two dimensional
transformation. The result is very good, the CEV smoother is able to
accurately find the hyperbolic profile.}
\label{fig:general}
\end{center}
\end{figure}

As an example, the relation $U_0=U_1\cdot U_2$
(Fig.~\ref{fig:general}a) is used.  Ten thousand equally distributed
data points in the range $(-1,1)$ for $U_1,\,U_2$ have been generated,
from these $U_0$ is obtained by multiplication. Gaussian,
$N(0,0.1^2$, noise has been added afterwards to each of the three
variables as noise realizations.  The estimated function coincides
well with the expectation (Fig.~\ref{fig:general}b). Note that in this
case ACE yields as a result $\log U_0 = \log U_1 + \log U_2$ because
of the additive structure of the model (\ref{eq:ace})

For this model, many applications with quasi-two-dimensional setting
can be found, e.g., in geological and hydrodynamical systems (e.g. the two
dimensional Euler equations).

\section{Application to spatiotemporal data}
\label{sec:apps}
Typical spatiotemporal models of interest for a physicist are coupled
map (CML) systems , coupled ordinary differential equations (ODE's) or
partial differential equations (PDE's). In all these models,
derivatives or finite differences are involved.  Here, the measured
quantity shall consist of a space-time record, given by a movie,
i.e. a series of consecutive pictures. From the time series in one
point one builds finite differences in time, from the pictures one
calculates differences in space.  This procedure is numerically not
trivial: addition (or subtraction, respectively) is numerically bad
conditioned and can yield extinction of digits, measurement noise is
then shifted to the first digits.  Basically, one must use the
apparatus known from numerical integration
\cite{NR,Lambert-73,Mitchell-Griffiths-80} to use correct sampling in
space and time. In the case of a periodic setup, accuracy loss can be
limited by using spectral methods \cite{NR,Canuto-etal-88}.

As an example, let us take the Complex Ginzburg-Landau Equation.
It describes a well pattern formating system above onset
\cite{Cross-Hohenberg-93}:
\begin{equation}
\partial_t A = (1+i c_1) \Delta A + (1- i c_3) f_1(|A|^2)A \;.
\end{equation}
The function $f_1$ is a complex nonlinear function. 
The task of data analysis -supposed that $A$ can be measured- is to
determine the nonlinearity $f_1$. To apply the ideas of backfitting we put
$U_0(x,t)=\partial_t A/A $, $U_1(x,t)=\Delta A/A$,
$U_2(x,t)=|A|^2$. Real and imaginary part must be treated separately.
Using the iteration procedure, one finally obtains estimates for the
parameters $c_1$, $c_3$ and the full function $f_1$. In a slightly
more complicated setup, this has been performed in
\cite{Voss-Kolodner-Abel-Kurths-99}. In fact, for this example the
numerically critical step is rather the correct evaluation of
derivatives than the application of backfitting.

If one wants to find a model, but there are no physical arguments
which derivatives should occur or it is even unclear if the measured
quantities are state variables or have undergone some measurement
transformations, one can still try to find a generalized additive
model. In that case, it can be worth to fit a discrete
model (CML), that represents the system dynamics,  avoiding a
good part of the trouble with finite differences (one can choose
mapping units big enough to avoid numerical problems).

\begin{figure}[htb]
a)\begin{center}
\includegraphics[clip,angle=0,width=0.8\textwidth]{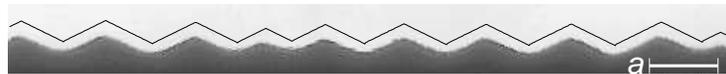}
\end{center}
b)\begin{center}
\includegraphics[clip,angle=0,width=0.8\textwidth]{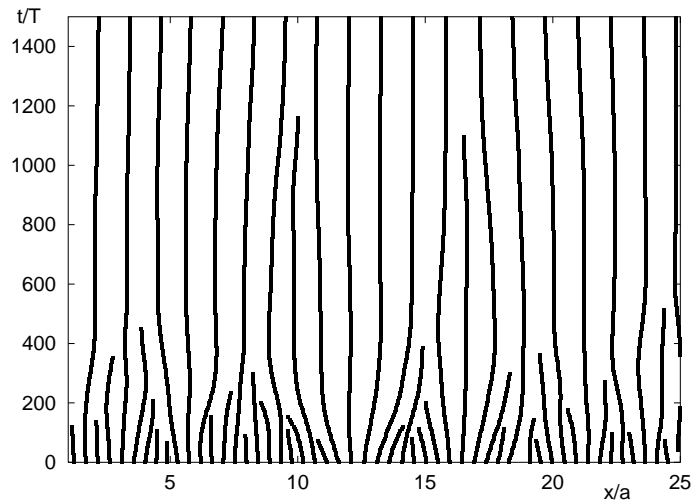}
\caption{a) Ripple profile as observed in the experiment. The driving
amplitude $a$ is marked. With the bottom at rest, water is oscillating
forth and back over the crests. Triangles (top, black line) are fitted
to raw picture (lower, dark profile) to determine the crests. b) The
experimental evolution of the position of the ripple crests starting
from ripples with a small length yields the shown space time plot.}
\label{fig:profile}
\end{center}
\end{figure}

A recently presented experiment considers underwater vortex ripples
\cite{Andersen-Chabanol-Hecke-01,AAKESU-2002}.  The ripple formation
is driven by a periodical motion of water over a sand bed with
given amplitude and frequency. A picture of the ripples -viewed from
the side- is given in Fig.~\ref{fig:profile} a), the resulting space
time plot is shown in Fig.~\ref{fig:profile} b). In this paper,
experimental details are not given, for those it is referred to the
literature \cite{AAKESU-2002,HHHEABS-01}. But the treatment of the
data and subsequent results are presented to demonstrate application
of the nonparametric regression.
The model is formed by coarse graining of space and time.  In time,
one uses a mapping from period to period, in space, ripples are
characterized by their length $\lambda$. The interaction between
ripples is modeled by a nonlinear interaction function which must be
concave by stability analysis.  The model is obtained for ripples
close to the final length; without discussing further the physical
mechanisms, it reads:
\begin{equation}
\label{eq:ripple}
\lambda_i(T+1) - \lambda_i(T) = - f(\lambda_{i-1(T)}) +f(\lambda_{i}(T)) -f(\lambda_{i+1}(T)) +\epsilon\;.
\end{equation}
The subscript $i$ denotes the space index. 

The process of data analysis involved the following steps: \\
i) calculating the ripple length, this has been done by fitting triangles
to the raw profiles and calculating the distances from crest to crest
(cf. Fig.~\ref{fig:profile}. The accuracy amounted to two digits.\\
ii) Calculation of the rhs of Eq.~(\ref{eq:ripple}), this has been
done by first calculating the difference, the result has been very
noisy due to the numerical extinction of digits. To compensate for
this effect, the differences have been smoothed and small numbers have
been discarded. 
Please note that no ``unsuitable'' points have been excluded (a
common, but often wrong technique for outliers), rather it is a {\it
numerical} requirement.  In any case, the model has been
developed from considerations of long ripples not too far from the
final length and thus very small ripples are not expected to follow
the model dynamics.\\
iii) To apply a backfitting algorithm, one has to identify the variables
$U_i$ as they have been denoted in the previous sections.  We put
formally $\Delta_t~\lambda~=~\lambda_i(T+1)-\lambda_i(T)=U_0$,
$\lambda_{i-1}(T)=U_1$, $\lambda_{i}(T)=U_2$, and $\lambda_{i+1}(T)=U_3$.  The
lhs shall be linear, so the model (\ref{eq:gam}) has been used.  For
this example, the functions are denoted by $f_1=f_l$, $f_2=f_c$,
$f_3=f_r$, for the left, center and right one, respectively. In the
present context this allows easier interpretation in terms of the
spatial structure. The data under consideration have been obtained by
nine experimental runs, each between 1000 and 6000 data points.

\begin{figure}
\begin{center}
\includegraphics[angle=0,width=1.0\textwidth]{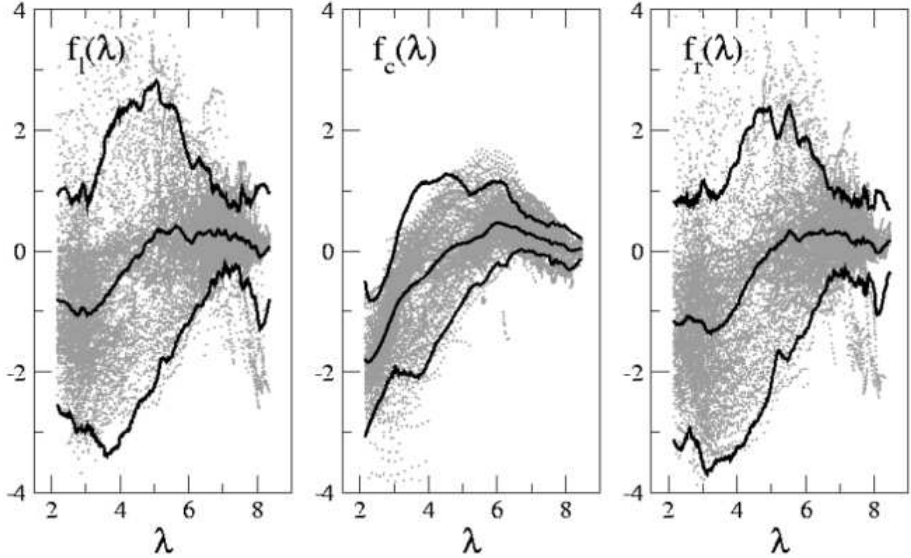}
\caption{Plot of the functions $f_l,\,f_c,\,f_r$ with pointwise
standard deviations (black lines) on top of the partial residuals
(grey dots). The left and right functions show wide scatter, reflected
in the individual correlations (numbers are given in the text). The curves
display the expected convex shape. For small ripple length there is a
systematic deviation in the left and right function.}
\label{fig:rip.whole}
\end{center}
\end{figure}

The experimental runs have been stationary as far as experimental
technique is concerned. Each data point $(\Delta_t
\lambda_i(T),\lambda_{i-1}(T),\lambda_{i}(T),\lambda_{i+1}(T))$ can be
considered as an independent realization of the underlying process.
Thus, the data from all nine measurements have been analyzed together.
The result is shown in Fig.~\ref{fig:rip.whole}.  For each function,
the importance criterion is calculated with $\Psi_0=0.732$,
$\Psi_1=0.560$, $\Psi_2=0.8604$ and $\Psi_3=0.6376$.  From these
numbers one obtains basically two informations: 1) There is a large
part of variance which cannot be modeled.  This can have two reasons,
the first is measurement noise, which has been reduced as much as
possible, the second is that parts of the dynamical equations (higher
order interactions or similar) are not included in the model. A
priori it is not easy (or even impossible) to distinguish which is the
reason, again the error in variables problem has to be treated with
care.  2) The growth of a ripple depends mainly on the length of this
ripple, but the neighbors cannot be neglected. This is seen by the
lower correlations for the neighboring functions.
 
As a basic test, the result functions must show the expected concave
shape. This is perfectly true in the range the model should hold.
There are, however differences between the neighbor functions
$f_l$, $f_r$ and the center one, $f_c$. Mainly one notes a very large
standard deviation at certain regions of $f_l,\, f_r$ and a systematic
deviation, a bend, at small ripple lengths. At this point the data
analysis result gives feedback to the experiment and the theoretical
modeling. It must be clarified if these differences have a physical
background or if they are artificial effects due to the measurement
procedure. This is ongoing work.

\begin{figure}
\begin{center}
\includegraphics[clip,angle=0,width=1.0\textwidth]{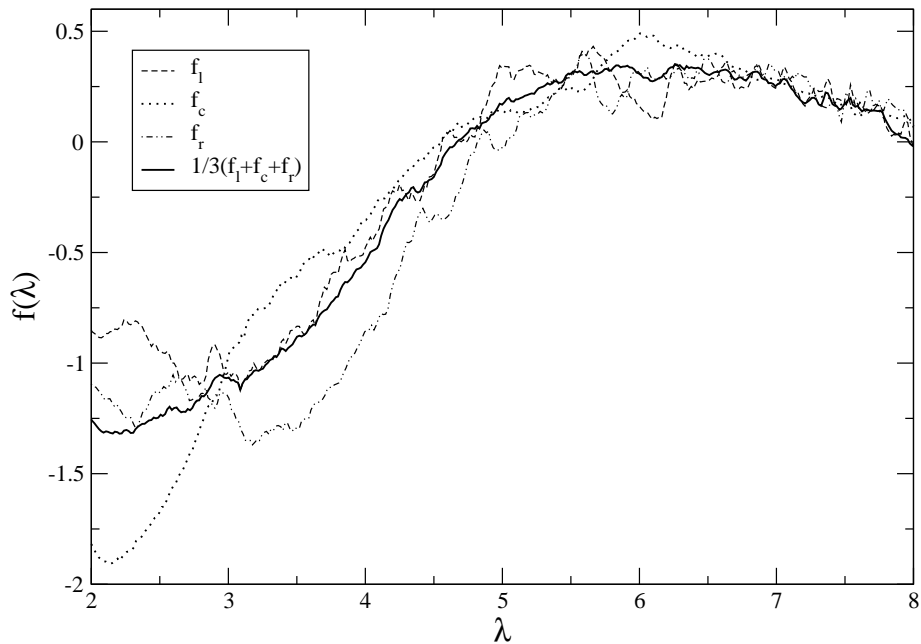}
\caption{Results from averaging over different runs
($f_l,f_c,f_r$). For long ripple lengths -the regime,  the model has been
developed for- one observes a very good coincidence. Theoretically, a
difference between sides and center is not expected; assuming them to
be equal, averaging yields the black line as an estimation for the true
interaction function.}
\label{fig:fin}
\end{center}
\end{figure}

In Fig.~\ref{fig:fin} the left, center and right function are plotted
together. Encouragingly, with respect to the established model, the
functions fall close together except for small ripple lengths. The
model has been developed in the region close to the final length, thus
the result confirms the model.  For small lengths, the result should
be considered as input for further modeling. In the case of small
ripple lengths, the deviations of the side functions to the center
indicate that in the small ripple length region, a difference between
center and sides exists. 
If one averages over left, center and right function, assuming that all
three functions are equal and differences arise from measurement errors
one obtains the thick black line in Fig.~\ref{fig:fin}. 

The model has been obtained as a great simplification of the highly
complex granular-fluid system. Given additionally a certain amount of
measurement and numerical errors from data processing, one can say
that the model is well confirmed by direct, nonparametric data
analysis.  Further activity aims at a more general modeling in the
sense of Eq.~(\ref{eq:general}) to check if there are differences in
the interaction functions of the sides and the center which is
theoretically very improbable. The direct justification of the model
(\ref{eq:ripple}) is a success not only for the modeler but as well
for statistical data analysis in physical experiments.

\section{Discussion and conclusion}
\label{sec:conclusion}
Some of the facts about generalized additive models, backfitting
algorithms and ACE have been presented.  With a step by step
explanation the basic ideas of algorithms and some technical details
have been discussed. In the examples, problems with nonlinear
transformation of data have been pointed out and the effect of
additional, uncorrelated variables has been investigated.  With data
from an example for a complex pattern forming system, it has been
shown how to apply the algorithms to spatio-temporal data and
explained the main difficulties for applications. With data analysis a
simple model has been validated, which has been developed by physical
arguments and coarse graining of the highly complex system of
underwater vortex ripples.
 
Generalized additive models occur often in physical and
engineering applications. The idea to obtain models directly from data
can be realized with the techniques developed in statistical science.
The focus of the presented methods lies in the possibility to
determine nonlinearities. Results from the nonparametric data analysis
help the theoretician to improve analytical models as well as the
applied physicist or engineer who just needs a working model to
represent the system dynamics.  These aspects can be closed to a
circuit - measurement - data analysis - theoretical modeling - which
can be iterated until a consistent model for the system under
consideration is obtained. It can be expected that the use of
nonparametric methods is helpful as well in situations where little is
known about the system. There, working models can be inferred without
further theoretical justification but as tools

While in this article additive models have been, there exist
techniques to solve for either additive or multiplicative models, e.g.
the marginal integration \cite{Linton-Nielsen-95} or alternatively
Breimans $\Pi$-method \cite{Breiman-91} (without a claim for
completeness). Implementations of the backfitting procedure or the ACE
algorithm exist for the S-PLUS language or the public domain clone
R-PLUS, or in the worldwide web \cite{code}.

Statistical approaches are promising tools for the analysis of complex
systems, since nonlinearities are automatically fetched.  Bayesian
formulations promise to yield better error models and another, future
application can be stochastic modeling, finding a representation of
the residuals in form of a noise term with measured probability
distribution.

\section{Acknowledgements}
I thank K. Andersen, J.~Kurths, A.~Pikovsky and A.~Politi for support
and discussion, a special thanks to H. Voss who led my attention to
this kind of problems. F. Schmidt has provided computer
wisdom, L.~Barker-Exp. was a constant source of inspiration.
I acknowledge support by the DFG (German research foundation), project
number PI 200/7-01.


\begin{thebibliography}{10}

\bibitem{Hastie-Tibshirani-86}
T.~Hastie and R.~Tibshirani.
\newblock Generalized additive models.
\newblock {\em Stat. Sci.}, 1:295--318, 1986.

\bibitem{Hastie-Tibshirani-90}
T.J. Hastie and R.J. Tibshirani.
\newblock {\em Generalized Additive Models}.
\newblock Chapman and Hall, London, 1990.

\bibitem{Kantz-Schreiber-97}
H.~Kantz and T.~Schreiber.
\newblock {\em Nonlinear time series analysis}.
\newblock Cambridge University Press, 1997.

\bibitem{Ott-94}
E.~Ott, T.~Sauer, and J.A. Yorke.
\newblock {\em Coping with Chaos}.
\newblock Series in Nonlinear Science. Wiley, New York, 1994.

\bibitem{NR}
B~P.~Flannery S.~A.~Teukolsky, W.~T.~Vetterling.
\newblock {\em Numerical Recipes in C: The Art of Scientific Computing}.
\newblock Cambridge University Press, Cambridge, 2nd edition, 1993.

\bibitem{Montgomery-92}
D.~C. Montgomery.
\newblock {\em Introduction to linear regression analysis}.
\newblock Wiley, New York, 1992.

\bibitem{Haerdle-Hall-93}
W.~H\"ardle and P.~Hall.
\newblock {\em On the backfitting algorithm for additive regression models}.
\newblock Cambridge University Press, Cambrige, 1993.

\bibitem{Buja-Hastie-Tibshirani-89}
A.~Buja, T.J. Hastie, and R.J. Tibshirani.
\newblock Linear smoothers and additive models.
\newblock {\em Ann. Stat.}, 17(2):453--555, 1989.

\bibitem{BF-85}
L.~Breiman and J.H. Friedman.
\newblock Estimating optimal transformations for multiple regression and
  correlation.
\newblock {\em J. Am. Stat. Assoc.}, 80:580--598, 1985.

\bibitem{Honerkamp-94}
J.~Honerkamp.
\newblock {\em Stochastic dynamical systems}.
\newblock VCH, New York, 1994.

\bibitem{Fan-Truong-93}
J.~Fan and Y.K. Truong.
\newblock Nonparametric regression with errors in variables.
\newblock {\em Ann. Stat.}, 21(4):1900--1925, 1993.

\bibitem{Carroll-Maca-Ruppert-99}
R.J. Carroll, J.D. Maca, and D.~Ruppert.
\newblock Nonparametric regression in the presence of measurement error.
\newblock {\em Biometrika}, 86(3):541--554, 1999.

\bibitem{Mammen-99}
E.~Mammen, O.~Linton, and J.~Nielsen.
\newblock The existence and asymptotic properties of a backfitting projection
  algorithm under weak conditions.
\newblock {\em Ann. Statist.}, 27(5):1443--1490, 1999.

\bibitem{Hastie-Tibshirani-00}
T.~Hastie and R.~Tibshirani.
\newblock Bayesian backfitting.
\newblock {\em Stat. Sci.}, 15(3):196--223, 2000.

\bibitem{Buja-90}
A.~Buja.
\newblock Remarks on functional canonical variates, alternating least squares
  methods and ACE.
\newblock {\em Ann. Stat.}, 18(3):1032--1069, 1990.

\bibitem{Renyi-59}
A.~R\'{e}nyi.
\newblock On measures of dependence.
\newblock {\em Acta.\ Math.\ Acad.\ Sci.\ Hungar.}, 10:441--451, 1959.

\bibitem{Bell-62}
C.B. Bell.
\newblock Mutual information and maximal correlation as measures of dependence.
\newblock {\em Ann.\ Math.\ Stat.}, 33:587--595, 1962.

\bibitem{Csaki-63}
P.~Csaki and J.~Fischer.
\newblock On the general notion of maximal correlation.
\newblock {\em Magyar Tud.\ Akad.\ Mat.\ Kutato Int.\ Kozl.}, 8:27--51, 1963.

\bibitem{Renyi-70}
A.~R\'{e}nyi.
\newblock Probability theory.
\newblock {\em Akad\'{e}miai Kiad\'{o}, Budapest}, 1970.

\bibitem{Lambert-73}
J.~Lambert.
\newblock {\em Computational Methods in Ordinary Differential Equations}.
\newblock Springer, New York, 1973.

\bibitem{Mitchell-Griffiths-80}
P.J. Mitchell and D.F. Griffith.
\newblock {\em The finite difference method in partial differential equations}.
\newblock Wiley, New York, 1980.

\bibitem{Canuto-etal-88}
C.~Canuto, M.Y. Hussaini, A.~Quarteroni, and T.A. Zang.
\newblock {\em Spectral methods in fluid dynamics}.
\newblock Springer, New York, 1988.

\bibitem{Cross-Hohenberg-93}
M.C. Cross and P.C. Hohenberg.
\newblock Pattern formation outside equilibrium.
\newblock {\em Rev. Mod. Phys.}, 65:3, 851-1112.

\bibitem{Voss-Kolodner-Abel-Kurths-99}
M.~Abel H.~Voss, P.~Kolodner and J.~Kurths.
\newblock Amplitude equations from spatiotemporal binary-fluid convection data.
\newblock {\em Phys. Rev. Lett.}, 83(17):3422--3425, 1999.

\bibitem{Andersen-Chabanol-Hecke-01}
K.H. Andersen, M.L. Chabanol, and M.~v.~Hecke.
\newblock Dynamical models for sand ripples beneathe surface waves.
\newblock {\em Phys Rev. E}, 63(6):66308, 1999.

\bibitem{AAKESU-2002}
K.H. Andersen, M.~Abel, J.~Krug, C.~Ellegaard, L.R. Soendergaard, and
  J.~Udesen.
\newblock Pattern dynamics of vortex ripples in sand: Nonlinear modeling and
  experimental validation.
\newblock {\em Phys. Rev. Lett.}, 88(23):4302, 2002.

\bibitem{HHHEABS-01}
J.~L. Hansen, M.~van Hecke, A.~Haaning, C.~Ellegaard, K.~H. Andersen, T.~Bohr,
  and T.~Sams.
\newblock Instabilities in sand ripples.
\newblock {\em Nature}, 410:324, 2001.

\bibitem{Linton-Nielsen-95}
O.~Linton and J.P. Nielsen.
\newblock A kernel method of estimating structured nonparametric regression
  based on marginal integration.
\newblock {\em Biometrika}, 82(1):93--100, 1995.

\bibitem{Breiman-91}
L.~Breiman.
\newblock The $\pi$ method for estimating multivariate functions from noisy
  data.
\newblock {\em Technometrics}, 33(2):125--143, 1991.

\bibitem{code}
http://www.gnu.org/directory/Mathematics/Statistics,\\http://www.splus.com,\\http://www.stat.physik.uni-potsdam.de/$\tilde{ }\,$markus/download,.

\end{thebibliography}
\end{document}